\RequirePackage{fix-cm}
\documentclass[smallcondensed]{svjour3}     
\smartqed  
\begin{document}

\title{Two-oscillator Kantowski-Sachs model of the Schwarzschild black hole interior
}

\titlerunning{Two-oscillator Kantowski-Sachs model of the Schwarzschild black hole interior}        

\author{Goran S. Djordjevic \and Ljubisa Nesic \and Darko Radovancevic 
}

\authorrunning{G. S. Djordjevic, Lj. Nesic, D. Radovancevic} 

\institute{Goran S. Djordjevic \at
              Faculty of Science and Mathematics, University of Nis, P.O. Box 224, 18000 Nis, Serbia\\
              Tel.: +381-18-533-015\\
              Fax: +381-18-533-014\\
              \email{gorandj@junis.ni.ac.rs}           
           \and
           Ljubisa Nesic \at
              Faculty of Science and Mathematics, University of Nis, P.O. Box 224, 18000 Nis, Serbia\\
              Tel.: +381-18-533-015\\
              Fax: +381-18-533-014\\
              \email{nesiclj@junis.ni.ac.rs}
            \and
            Darko Radovancevic \at
            Faculty of Science and Mathematics, University of Nis, P.O. Box 224, 18000 Nis, Serbia\\
            Tel.: +381-18-533-015\\
            Fax: +381-18-533-014\\ 
            \email{darko.radovancevic@pmf.edu.rs}
}


\maketitle

\begin{abstract}
In this paper the interior of the Schwarzschild black hole, which is presented as a vacuum homogeneous and anisotropic Kantowski-Sachs minisuperspace cosmological model, is considered. Lagrangian of the model is reduced by a suitable coordinate transformation to Lagrangian of two decoupled oscillators with the same frequencies and with zero energy in total (an oscillator-ghost-oscillator system). The model will be presented in a classical, a $p$-adic and a noncommutative case. Then, within the standard quantum approach Wheeler-DeWitt equation and its general solutions, i.e. a wave function of the model, will be written, and then an adelic wave function will be constructed. Finally, thermodynamics of the model will be studied by using the Feynman-Hibbs procedure.



\end{abstract}

\section{Introduction}
The study of physics of the black holes takes an important place in the theory of quantum gravity. Namely, the general relativity along with the initial singularity of the universe predicts the existence of local singularities within black holes. That enables us to come to a significant conclusion related to the early stages of the creation of the universe by studying the characteristics of these objects. In both cases, the dynamics of the matter and space-time at the Planck scale demands quantum approach. In the absence of a complete theory of quantum gravity, it is important to analyse particular classical cosmological models and their quantum versions within the approach that is using canonical quantization or path integral quantization. Both of these approaches demand the determination of the wave function (of the universe) for a quantum cosmological model. By doing so in the first case (canonical qantization) the wave function is acquired through solving Wheeler-DeWitt equation \cite{jpap1}, in the second case (path integral quantization) it can be done by Feynman minisuperspace propagator with appropriate boundary condition (the best known are the Hartle-Hawking no-boundary proposal \cite{jpap3} and the Vilenkin tunneling proposal \cite{jpap2}).

This paper presents the Schwarzschild black hole interior as a homogeneous and anisotropic Kantowski-Sachs minisuperspace cosmological model without a matter field and the cosmological constant.This approach is based on a diffeomorphism between the Schwarzschild solution of the Einstein field equation and the corresponding solution of this cosmological model. Namely, within the Schwarzschild sphere, the coordinate transformation $t\leftrightarrow r$, transforms Schwarzschild metric to anisotropic Kantowski-Sachs metric. Within the classical analytical approach, Lagrangian of the model, by a suitable coordinate transformation will become Lagrangian of an oscillator-ghost-oscillator system of two decoupled oscillators, with equal frequencies and with zero energy in total.

Motivated by possible non-Archimedean and/or noncommutative structure of the space-time at the Planck scale \cite{autbk1}, we will discuss a $p$-adic and then a noncommutative form of this model. The prediction, of the space-time discreetness structure around Planck distances ($\leq 10^{-33}$ cm), is common to both of these approaches ($p$-adic and noncommutative). In $p$-adic case this discreetness is implicitly present \cite{jpap{13}}, in noncommutative case it is explicitly present through commutation relations of noncommutative coordinates and/or their conjugate momenta. Noncommutative coordinates are used for the first time by Wigner \cite{jpap{15}} and Snyder \cite{jpap{16}}. This idea, used by Connes \cite{jpap{17}} and Woronowicz \cite{jpap{18}} in noncommutative geometry, enabled the development of the new formulation of quantum gravity through noncommutative differential calculus \cite{jpap{188},seri{19},jpap{20}}.

After determination and decoupling of Langrangian for the classical anisotropic Kantowski-Sachs 
model of the Schwarzschild black hole interior, for $p$-adic form of the model, we will determine 
the propagator and the conditions of the vacuum states existence. In the noncommutative case, 
classical action for the two types of the Euler-Lagrange equation solutions, will be determined 
from noncommutative Lagrangian. After that, we will write Wheeler-DeWitt equation and determine 
its general solution i.e. the wave function which corresponds to the interior of a non-rotating 
and non-charged (Schwarzschild) black hole, according to Hamiltonian formalism for the quantum 
form of the model in commutative case. Then, an adelic wave function of the model will be constructed 
as well. At the end of the paper, Hawking temperature and entropy with logaritmic quantum corrections
will be determined by the application of the Feynman-Hibbs procedure on the Wheeler-DeWitt equation of the model.

\section{The Classical Model}

In canonical formulation of the general relativity the starting point is 3+1 decomposition of metric (in natural units, where $\hbar=c=1$):
\begin{equation}
ds^2=g_{\mu\nu}dx^{\mu}dx^{\nu}=-N^2dt^2+h_{ik}(dx^{i}+N^{i}dt)(dx^{k}+N^{k}dt),
\end{equation}
where $N$ is so-called the lapse function, and $N^i$ are commponents of the shift vector. The Einstein-Hilbert action, within this approach, has the following form:
\begin{eqnarray}
&S&=S_g+S_{YGH}+S_m
\nonumber\\
&=&\frac{1}{16\pi G}\!\int_{M}\!\left[^{(3)}R\!+\!K^{ik}K_{ik}\!-\!K^2\!-\!2\Lambda\right] N\sqrt{\zeta}\,dt\,d^{3}x\!+\!\int_{M}\!L_{m}N\sqrt{\zeta}\,dt\,d^{3}x,
\end{eqnarray}
where $S_g$ is the gravitational field action, $S_{YGH}$ is the York-Gibbons-Hawking boundary term, $S_m$ is the action for the Lagrangian $L_m$ of the matter, $G$ is the gravitational constant, $^{(3)}R$ is the Ricci scalar of the intrinsic 3-geometry, $\zeta$ is the determinant of the intrinsic 3-metric tensor $h_{ik}$ (or first fundamental form), $K_{ik}$ is the extrinsic curvature tensor (or second fundamental form), $\Lambda$ is the cosmological constant and $K=K^{i}{}_{i}$.

Starting with the Schwarzschild metric element for centrally symmetric gravitational field:
\begin{equation}
ds^2=-(1-\frac{r_g}{r})dt^2+(1-\frac{r_g}{r})^{-1}dr^2+r^2(d\theta^2+\sin^2\theta d\varphi^2),
\end{equation}
which diverges in case $r=0$ (Schwarzschild or gravitational singularity) and on so-called Schwarzschild sphere (the event horizon) for $r=r_g$, where $r_g=\frac{2Gm}{c^2}$ is the gravitational or Schwarzschild radius for the mass $m$. This other singularity is ''apparent'' or coordinate singularity and can be removed by a suitable coordinate transformation (for example Kruskal-Szekeres transformation).

On the other hand, we can see from (3) that when $r < r_g$, i.e. inside of the event horizon, Schwarzschild metric components $g_{00}=g_{tt}$ and $g_{11}=g_{rr}$ changed their signs. This suggests that space and time change roles in a way, once the event horizon is passed. Namely, the time coordinate, for an observer outside the event horizon, becomes spatial-like for another observer inside the Schwarzschild sphere. In this sense, for both observers, there is one possible direction of movement, for an outside observer in time and for an inside observer in space (right to singularity). In regard to what was previously said, we can consider a possibility that Schwarzschild sphere interior could be described by a square metric form acquired by substitution $t\leftrightarrow r$ in its standard spherical form (3), i.e. form:
\begin{equation}
ds^2=-(\frac{r_g}{t}-1)^{-1}dt^2+(\frac{r_g}{t}-1)dr^2+t^2(d\theta^2+\sin^2\theta d\varphi^2),
\end{equation}
which metric components explicitly depend on time. The previous equation is of the following form:
\begin{equation}
ds^2=-\frac{N^2(t)}{\nu{(t)}} dt^2+\nu{(t)}dr^2+h^2(t)(d\theta^2+\sin^2\theta d\varphi^2),
\end{equation}
which is actually a homogeneous and anisotropic Kantowski-Sachs metric form with two scale factors $h$ and $\nu$. In this case we can see from (4) i (5) that $\nu{(t)}=\frac{r_g}{t}-1$, $h(t)=t$ and $N(t)=1$.

From the Einstein-Hilbert action (2), without a matter field ($L_m$=0) and the cosmological constant ($\Lambda =0$) with the metric (5) one gets Lagrangian for the vacuum Kantowski-Sachs model of the interior of the Schwarzschild black hole in the form \cite{jpap{21}}:
\begin{equation}
L=-\frac{V_0}{8\pi G}\left[\frac{1}{N}(h\dot{h}\dot{\nu}+{\dot{h}}^2\nu)-N\right],
\end{equation}
where $V_0=4\pi\int\,dr$ is a volume (in the unit of the length) of the space part of the action which is treated to be a finite. By the rescaling of the lapse function:
\begin{equation}
N=\sqrt{h^2\nu}\tilde{N},
\end{equation}
with transformations \cite{jpap{22}}:
\begin{equation}
h=\frac{1}{4}(u-v)^2, \quad \nu=\left[\frac{u+v}{u-v}\right]^2,
\end{equation}
Lagrangian (6) becomes decoupled:
\begin{eqnarray}
L&=&-M_{pl}^2{V_0}{\tilde{N}}^{-1}({\dot{u}}^2-{\dot{v}}^2)+M_{pl}^2{V_0}\tilde{N}(u^2-v^2)
\nonumber\\
&=&-\frac{M_{pl}^2{V_0}}{\tilde{N}}\left[({\dot{u}}^2-{\tilde{N}}^{2}u^2)-({\dot{v}}^2-{\tilde{N}}^{2}v^2)\right],
\end{eqnarray}
where $M_{pl}=\sqrt{\frac{{\hbar}c}{8{\pi}G}}\approx 2.43\times10^{18}$ GeV is the reduced Planck mass. Momenta conjugate to coordinates $u$, $v$, $\tilde{N}$ are, respectively:
\begin{eqnarray}
\pi_u&=&\frac{\partial{L}}{\partial{\dot{u}}}=-\frac{2{V_0}M_{pl}^2}{\tilde{N}}\dot{u},\\
\pi_v&=&\frac{\partial{L}}{\partial{\dot{v}}}=\frac{2{V_0}M_{pl}^2}{\tilde{N}}\dot{v},\\
\pi_{\tilde{N}}&=&\frac{\partial{L}}{\partial{\dot{\tilde{N}}}}=0.
\end{eqnarray}
The last formula represents so-called the primary constraint. Hamiltonian of the system within canonical variables is:
\begin{eqnarray}
H&=&\dot{u}\pi_u+\dot{v}\pi_v+\dot{\tilde{N}}\pi_{\tilde{N}}-L
\nonumber\\
&=&-\frac{\tilde{N}}{4M_{pl}^2{V_0}}(\pi_u^2-\pi_v^2)-{V_0}M_{pl}^2\tilde{N}(u^2-v^2).
\end{eqnarray}
Considering the fact that there is the primary constraint (12), Hessian of the system is equal to zero, which means that Lagrangian of the system is a singular one i.e. that the system of dynamic differential equations of second order by $u$, $v$, $\tilde{N}$ does not have an unambiguous solution by the highest order derivatives of these functions. This implies the absence of an unambiguous particular solution for the given starting conditions. Then, the Hamiltonian of the system is not a unique one. In that case, the total Hamiltonian is introduced:
\begin{equation}
H_T=H+\lambda\pi_{\tilde{N}}=-\frac{\tilde{N}}{4M_{pl}^2{V_0}}(\pi_u^2-\pi_v^2)-{V_0}M_{pl}^2\tilde{N}(u^2-v^2)+\lambda\pi_{\tilde{N}},
\end{equation}
where $\lambda=\lambda{(t)}$ is a Lagrange multiplier. The secondary (Hamiltonian) constraint is in that case:
\begin{equation}
\dot\pi_{\tilde{N}}=\{{\pi_{\tilde{N}}},H_T\}_{pz}=\{{\pi_{\tilde{N}}},H\}_{pz}=-\frac{\partial{H}}{\partial{\tilde{N}}}={\cal{H}}=0,
\end{equation}
where:
\begin{equation}
{\cal{H}}=-\frac{1}{4M_{pl}^2{V_0}}(\pi_u^2-\pi_v^2)-{V_0}M_{pl}^2(u^2-v^2),
\end{equation}
which will result in Wheeler-DeWitt equation in quantisation procedure.

Because of the possibility of a choice of the lapse function $\tilde{N}$ (gauge fixing), we can consider factor $-\frac{M_{pl}^2{V_0}}{\tilde{N}}$ in formula (9) as a constant. Considering $\tilde{N}=\omega$, and bearing in mind the invariance of the system dynamics in relation to multiplication of Lagrangian  with constant, instead of (9) we can consider:
\begin{equation}
L=({\dot{u}}^2-{\omega}^{2}u^2)-({\dot{v}}^2-{\omega}^{2}v^2),
\end{equation}
as a Lagrangian which describes the model. This is the Lagrangian of a oscillator-ghost-oscillator system with zero energy. The corresponding Euler-Lagrangian equations are:
\begin{equation}
\ddot{u}+\omega^2u=0, \quad \ddot{v}+\omega^2v=0.
\end{equation}
The general solution of (18) is:
\begin{equation}
u=A_1\cos(\omega{t}+A_2), \quad v=B_1\cos(\omega{t}+B_2),
\end{equation}
where $A_1, A_2, B_1, B_2$ are constants of integration. By substituting (19) into Hamiltonian constraint
(15), it is obtained that $A_1=\pm{B_1}$. 

Particular solution for initial conditions $u(t')=u'$, $u(t'')=u''$, $v(t')=v'$ and $v(t'')=v''$, is:
\begin{equation}
u(t)=u'\frac{\sin(\omega(t''-t))}{\sin(\omega(t''-t'))}+u''\frac{\sin(\omega(t-t'))}{\sin(\omega(t''-t'))},
\end{equation}
\begin{equation}
v(t)=v'\frac{\sin(\omega(t''-t))}{\sin(\omega(t''-t'))}+v''\frac{\sin(\omega(t-t'))}{\sin(\omega(t''-t'))}.
\end{equation}

By substituting (20) and (21) into (17) the classical Lagrangian $L^{cl}$ is obtained, and after its integration, in time, we get the classical action for this model:
\begin{eqnarray}
&S^{cl}&(u'',v'',t'';u',v',t')=\int_{t'}^{t''}L^{cl}\,dt
\nonumber\\
&=&\omega\left[\frac{{u''}^2+{u'}^2}{\tan(\omega T)}-\frac{2u''u'}{\sin(\omega T)}\right]-\omega\left[\frac{{v''}^2+{v'}^2}{\tan(\omega T)}-\frac{2v''v'}
{\sin(\omega T)}\right],
\end{eqnarray}
where $T=t''-t'$.

\section{The Classical and Quantum Model in the $p$-Adic Case}

In $p$-adic space-time the Lagrangian (17) and equations of motion in general have the same form, but with variables which take values from the field of $p$-adic numbers $Q_p$. By solving $p$-adic equations of motion for initial conditions $u(t')=u'$, $u(t'')=u''$, $v(t')=v'$ and $v(t'')=v''$ we obtain a particular $p$-adic valued solution. Then, in the same manner as in the standard case, we obtain a classical $p$-adic action, that has the same form as in the real case (22), but with $p$-adic values of variables and different domain of definition and convergence of $p$-adic function \cite{autbk1}. Bearing this in mind, we can write a classical $p$-adic action of the model in the following form:
\begin{eqnarray}
&S_p^{cl}&(u'',v'',t'';u',v',t')=\int_{t'}^{t''}L_p^{cl}\,dt
\nonumber\\
&=&\omega\left[\frac{{u''}^2+{u'}^2}{\tan(\omega T)}-\frac{2u''u'}{\sin(\omega T)}\right]-
\omega\left[\frac{{v''}^2+{v'}^2}{\tan(\omega T)}-\frac{2v''v'}
{\sin(\omega T)}\right],
\end{eqnarray}
where $T=t''-t'$, $t'$, $t''\in Q_p$. In this case, $\sin x $ and $\tan x $ are $p$-adic counterparts of the standard trigonometry functions that are defined as series (which have the same form as in the real case) whose domain of convergence is $G_p=\{x\in Q_p: |x|_p\leq|2p|_p\}$ \cite{autbk1}.
\newline
\indent Dynamics of a $p$-adic quantum model is described by a unitary evolution operator $U(t)$ given in an integral form:
\begin{eqnarray}
U_p(t)\psi (x)=\int_{Q_p}{\cal K}_t(x,y)\psi (y)\,dy,
\end{eqnarray}
where ${\cal K}_t(x,y)$ is a quantum-mechanical propagator which is defined by Feynman functional integral:
\begin{eqnarray}
{\cal K}_p(x'',t'';x';t')=\int_{x',t'}^{x'',t''}\chi_p\left(-\frac{1}{2\pi \hbar} \int_{t'}^{t''} L(\dot{q},q,t)\,dt\right)\,\textit{D}q,
\end{eqnarray}
where $\chi_p(a)=\exp(2\pi i \{a\}_p)$ is a $p$-adic additive character. Here, $\{a\}_p$ denotes the fractional part of $a\in Q_p$. The general formula for propagators for quadratic Lagrangians, valid in real, $p$-adic and adelic spaces was found \cite{jpap{24},jpap{xx}}. For any decoupled quadratic system with two degrees of freedom, the $p$-adic propagator is:
\begin{eqnarray}
{\cal K}_p(u'',v'',t'';u',v',t')&=&
\lambda_p\left(-\frac{1}{2}\frac{\partial^2{S_p^{cl}}}{\partial{u''}\partial{u'}}\right)
\lambda_p\left(-\frac{1}{2}\frac{\partial^2{S_p^{cl}}}
{\partial{v''}\partial{v'}}\right)
\nonumber\\
&\times &\left|\frac{\partial^2{S_p^{cl}}}{\partial{u''}\partial{u'}}\right|_p^\frac{1}{2}
\left|\frac{\partial^2{S_p^{cl}}}{\partial{v''}\partial{v'}}\right|_p^\frac{1}{2}
\chi_p(-S_p^{cl}),
\end{eqnarray}
where $\lambda_p(x)$ is an arithmetic complex-valued function \cite{autbk1}. Substituting (23) into (26) one gets:
\begin{equation}
{\cal K}_p(u'',v'',t'';u',v',t')=\left|\frac{2}{T}\right|_p
\chi_p(-S_p^{cl}),
\end{equation}
i.e. that is a $p$-adic propagator for this two-oscillator model we consider.

The existence of vacuum/ground states is of essential importance for every quantum-mechanical model. Conditions for the existence of vacuum $p$-adic states in form of (characteristic) $\Omega$ function are determined with \cite{jpap{13}}:
\begin{equation}
\int_{|u'\!|\le{p^{\!-\!\nu}}}\!\int_{|v'\!|\le{p^{\!-\!\mu}}}\!
{\cal K}_p(u'',v'',t'';u',v',t') \,du' \,dv'\!=\!\Omega(p^\nu|u''|_p)\Omega(p^\mu|v''|_p),
\end{equation}
($p$-adic function $\Omega(|x|_p)$ has the value 1 or 0, for $|x|_p\leq 1$ or $|x|_p>1$, respectively). By substituting (27) into (28) we obtain $p$-adic vacuum states:
\begin{equation}
\Psi_p^{(\nu,\mu)}(u,v)=\Omega(p^\nu|u|_p)\Omega(p^\mu|v|_p), \quad \nu,\mu=0,\pm 1,\pm2...
\end{equation}
that exist within the region of convergence, $G_p=\{\omega T\in Q_p: |\omega T|_p\leq|2p|_p\}$, of analytical $p$-adic functions $\sin(\omega T)$ i $\tan(\omega T)$.

\indent In general, $p$-adic vacuum state is a degenerated one \cite{jpap{13}}. For instance, vacuum states of $\delta$ type function satisfy \cite{jpap{13}}:
\begin{equation}
\int_{|u'\!|=p^{\nu}}\!\int_{|v'\!|=p^{\mu}}\! {\cal K}_p(u'',v'',t'';u',v',t') \,du' \,dv'\!=\!\delta(p^\nu\!-\!|u''|_p)\delta(p^\mu\!-\!|v''|_p).
\end{equation}
By substitution of (27) into (30) one gets:
\begin{equation}
\Psi_p^{(\nu,\mu)}(u,v) =
\left\{ \begin{array}{ll}
\delta(p^\nu-|u|_p)\delta(p^\mu-|v|_p), & \mbox{$|T|_p\leq p^{2\nu,\mu-2}$},\\
\delta(2^\nu-|u|_2)\delta(2^\mu-|v|_2), & \mbox{$|T|_2\leq 2^{2\nu,\mu-4}$},
\end{array} \right.
\end{equation}
for $\nu,\mu\ =0,-1,-2...$

It is worth emphasizing that these conditions, in (29) and (31), are very important in the context of $p$-adic and adelic formalism. Despite their physical meaning is not yet completely well understood it has a strong impact on physical constraints of the model on Archimedean and non-Archimedean spaces.   

\section{The Model on Noncommutative Space}

Space-time structure at the Planck scale is one of the most challenging question in high energy physics. Two of the most justified approaches to this problem are: non-Archimedean spaces, which we previously considered, and a noncommutative approach which will be presented in this part of the paper.

In the presence of noncommutativity of the spatial type i.e. $\lbrace u,v \rbrace_{pz}\!=\!\theta\!\not=\!0$,
$\lbrace u,\pi_{u} \rbrace_{pz}\!=\lbrace v,\pi_{v} \rbrace_{pz}\! =1$, $\lbrace u,\pi_{v} \rbrace_{pz}\! =\lbrace v,\pi_{u} \rbrace_{pz}\! =0$ and $\lbrace\pi_{u},\pi_{v} \rbrace_{pz}\!=\!0$, with transformations:
\begin{equation}
u\rightarrow u-\frac{\theta}{2}\pi_{v}=u+\theta\dot{v},
\end{equation}
\begin{equation}
v\rightarrow v+\frac{\theta}{2}\pi_{u}=v+\theta\dot{u},
\end{equation}
two-oscillator Lagrangian (17) becomes:
\begin{equation}
L_\theta=\left[(1+\omega^2\theta^2)\dot{u}^2-\omega^2u^2\right]-\left[(1+\omega^2\theta^2)\dot{v}^2-\omega^2v^2\right]
+2\theta\omega^2\left[\dot{u}v-\dot{v}u\right].
\end{equation}
The corresponding Euler-Lagrangian equations are:
\begin{equation}
\ddot{u}+2\theta\omega_\theta^2\dot{v}+\omega_\theta^2 u=0, \quad
\ddot{v}+2\theta\omega_\theta^2\dot{u}+\omega_\theta^2 v=0,
\end{equation}
where $\omega_\theta=\frac{\omega}{\sqrt{1+{\theta}^2{\omega}^2}}$.
In commutative regime ($\theta =0$) Lagrangian (34) and equations (35) becomes (17) and (18), respectively.

\indent On the other hand, the general trigonometric solution of the sistem (35) is:
\begin{eqnarray}
u(t)&=&C_1\cos(\Omega_1t)+C_2\sin(\Omega_1t)+C_3\cos(\Omega_2t)+C_4\sin(\Omega_2t),\\
v(t)&=&-\frac{2\theta\omega_{\theta}^2\Omega_1}{\omega_{\theta}^2-\Omega_1^2}C_2\cos(\Omega_1t)+\frac{2\theta\omega_{\theta}^2\Omega_1}{\omega_{\theta}^2-\Omega_1^2}C_1\sin(\Omega_1t)
\nonumber\\
&{}&-\frac{2\theta\omega_{\theta}^2\Omega_2}{\omega_{\theta}^2-\Omega_2^2}C_4\cos(\Omega_2t)
+\frac{2\theta\omega_{\theta}^2\Omega_2}{\omega_{\theta}^2-\Omega_2^2}C_3\sin(\Omega_2t),
\end{eqnarray}
where:
\begin{equation}
\Omega_{1,2}=\left[\frac{(2\omega_\theta^2-4\theta^2\omega_\theta^4){\pm}\sqrt{(2\omega_\theta^2-4\theta^2\omega_\theta^4)^2-4\omega_\theta^4}}{2}\right]^\frac{1}{2},
\end{equation}
with the condition that $\omega_\theta^2\neq\Omega_{1,2}^2>0$. For initial conditions $u(0)=u'$, $u(T)=u''$, $v(0)=v'$ and $v(T)=v''$, from (36) and (37) we obtain a classical solution, whose replacement in (34) gives classical Lagrangian. We obtain the corresponding classical action in the standard manner in the interval $[0,T]$ in the noncommutative case:
\begin{eqnarray}
&S_\theta^{cl}&(u'',v'',T;u',v',0)=
\frac{1}{2}\gamma_{11}{u'}^2+\frac{1}{2}\gamma_{22}{u''}^2+\frac{1}{2}\gamma_{33}{v'}^2+\frac{1}{2}\gamma_{44}{v''}^2
\nonumber\\
&+& \gamma_{12}u'u''+\gamma_{13}u'v'+\gamma_{14}u'v''+
\gamma_{23}v'u''+\gamma_{24}u''v''+\gamma_{34}v'v'',
\end{eqnarray}
where:
\begin{eqnarray}
\gamma_{ij}&=&\frac{1}{2\Omega_1}[(\alpha_{2i}\alpha_{2j}-\alpha_{1i}\alpha_{1j})\sin(2\Omega_1T)
+(\alpha_{2i}\alpha_{1j}+\alpha_{2j}\alpha_{1i})(\cos(2\Omega_1T)-1)]K_1^{-}
\nonumber\\
&+&\frac{1}{2\Omega_2}[(\alpha_{4i}\alpha_{4j}-\alpha_{3i}\alpha_{3j})\sin(2\Omega_2T)
+(\alpha_{4i}\alpha_{3j}+\alpha_{4j}\alpha_{3i})(\cos(2\Omega_2T)-1)]K_2^{-}
\nonumber\\
&+&\frac{1}{\Omega_1+\Omega_2}[(\alpha_{2i}\alpha_{4j}+\alpha_{2j}\alpha_{4i}-\alpha_{1i}\alpha_{3j}-\alpha_{1j}\alpha_{3i})\sin((\Omega_1+\Omega_2)T)
\nonumber\\
&+&(\alpha_{1i}\alpha_{4j}+\alpha_{1j}\alpha_{4i}+\alpha_{2i}\alpha_{3j}+\alpha_{2j}\alpha_{3i})(\cos((\Omega_1+\Omega_2)T)-1)]K_3^{-}
\nonumber\\
&+&\frac{1}{\Omega_1-\Omega_2}[(\alpha_{2i}\alpha_{4j}+\alpha_{2j}\alpha_{4i}+\alpha_{1i}\alpha_{3j}+\alpha_{1j}\alpha_{3i})\sin((\Omega_1-\Omega_2)T)
\nonumber\\
&+& (\alpha_{1i}\alpha_{4j}+\alpha_{1j}\alpha_{4i}-\alpha_{2i}\alpha_{3j}-\alpha_{2j}\alpha_{3i})(\cos((\Omega_1-\Omega_2)T)-1)]K_3^{+}
\nonumber\\
&+&T(\alpha_{2i}\alpha_{2j}+\alpha_{1i}\alpha_{1j})K_1^{+}+T(\alpha_{4i}\alpha_{4j}+\alpha_{3i}\alpha_{3j})K_2^{+}, \quad i\leq j=1,2,3,4.
\end{eqnarray}
Coefficients $\alpha_{ij}$ and $K_i^{\pm}$ are given by (75) and (76), respectively, in Appendix A. Expressions (39) and (40) are formally of the same form as the expressions in Ref. \cite{jpap{226}}.

The general form of the Feynman kernel for quadratic action in two-dimensional noncommutative space is \cite{jpap{26}}:
\begin{eqnarray}
{\cal K}_\theta(u'',v'',T;u',v',0)&=&\frac{1}{2\pi i \hbar}
\left[det\left(
\begin{array}{cc}
{-\frac{\partial^2 S_\theta^{cl}}{\partial u''\partial u'}}
&{-\frac{\partial^2 S_\theta^{cl}}{\partial u''\partial v'}}\\
{-\frac{\partial^2 S_\theta^{cl}}{\partial v''\partial u'}}
&{-\frac{\partial^2 S_\theta^{cl}}{\partial v''\partial v'}}
\end{array}
\right)\right]^\frac{1}{2}
\nonumber\\
&\times&\exp\left(\frac{i}{\hbar}S_\theta^{cl}(u'',v'',T;u',v',0)\right).
\end{eqnarray}
Replacing (39) into (41) we obtain the propagator:
\begin{equation}
{\cal K}_\theta(\!u'',\!v'',\!T;\!u',v',\!0\!)\!=\!\frac{1}{2\pi i \hbar}
\sqrt{\gamma_{12}\gamma_{34}\!-\!\gamma_{14}\gamma_{23}}
\exp\!\left(\!\frac{i}{\hbar}S_\theta^{cl}(\!u'',\!v'',\!T;\!u',\!v',\!0\!)\!\right)\!.
\end{equation}

On the other hand, the system of equations (35) has a solution of the following type \cite{jpap{266}}, as well:
\begin{equation}
u(t)=A_+e^{\theta\omega^2t}\sin(\omega_{\theta}t+\delta_+)+A_-e^{-\theta\omega^2t}\sin(\omega_{\theta}t+\delta_-),
\end{equation}
\begin{equation}
v(t)=-A_+e^{\theta\omega^2t}\sin(\omega_{\theta}t+\delta_+)+A_-e^{-\theta\omega^2t}\sin(\omega_{\theta}t+\delta_-),
\end{equation}
where $\omega_{\theta}=\omega\sqrt{1-\theta^2\omega^2}$.
By substitution of the initial conditions $u(0)=u'$, $u(T)=u''$, $v(0)=v'$ and $v(T)=v''$ in (43) and (44) one gets constants of integration as follows:
\begin{equation}
A_{\pm}\!=\!\frac{[\!(u''\!\mp\!v'')\!^2\!+\!(u'\!\mp\!v')\!^2e^{{\pm2}\theta\omega^2T}\!-\!2(u''\!\mp\!v'')\!(u'\!\mp\!v')e^{\pm\theta\omega^2T}\cos(\!\omega_{\theta}T)]^{1/2}}{2e^{\pm\theta\omega^2T}\sin(\!\omega_{\theta}T\!)\!},
\end{equation}
\begin{equation}
\delta_{\pm}=\arcsin\left(\frac{u'{\mp}v'}{A_{\pm}}\right).
\end{equation}
By substitution (45) and (46) into (43) and (44) we obtain a classical solution of equations of motions whose replacement in (34) gives the classical Lagrangian of the model for the latter solution of Euler-Lagrange equations (35). Then, in the standard manner we obtain the corresponding classical action:
\begin{eqnarray}
&{}&S_\theta^{cl}(u'',v'',T;u',v',0)=(1-\frac{1}{2}\omega^2\theta^2-\omega^4\theta^4)\omega_{\theta}
\nonumber\\
&\times&\{\!A_+^2[e^{2\theta\omega^2T}\sin(2\omega_{\theta}T\!+\!2\delta_+\!)\!-\!\sin(2\delta_+\!)\!+\!\frac{\theta\omega^2}{\omega_{\theta}}\!(\!e^{2\theta\omega^2T}\cos(2\omega_{\theta}T\!+\!2\delta_+\!)\!-\!\cos(2\delta_+\!)\!)\!]
\nonumber\\
&+&A_-^2[e^{-2\theta\omega^2T}\!\sin(2\omega_{\theta}T\!+\!2\delta_-\!)\!-\!\sin(2\delta_-\!)
\nonumber\\
&-&\frac{\theta\omega^2}{\omega_{\theta}}\!(\!e^{-2\theta\omega^2T}\!\cos(2\omega_{\theta}T\!+\!2\delta_-\!)\!-\!\cos(2\delta_-\!)\!)\!]\}+\theta\omega^2(1-\omega^4\theta^4)
\nonumber\\
&\times&\{\!A_+^2[-e^{2\theta\omega^2T}\!\cos(2\omega_{\theta}T\!+\!2\delta_+\!)\!+\!\cos(2\delta_+\!)\!+\!\frac{\theta\omega^2}{\omega_{\theta}}\!(\!e^{2\theta\omega^2T}\!\sin(2\omega_{\theta}T\!+\!2\delta_+\!)\!-\!\sin(2\delta_+\!)\!)\!]
\nonumber\\
&+&A_-^2[e^{-2\theta\omega^2T}\!\cos(2\omega_{\theta}T\!+\!2\delta_-\!)\!-\!\cos(2\delta_-\!)
\nonumber\\
&+&\frac{\theta\omega^2}{\omega_{\theta}}\!(\!e^{-2\theta\omega^2T}\!\sin(2\omega_{\theta}T\!+\!2\delta_-\!)\!-\!\sin(2\delta_-\!)\!)\!]\}
\nonumber\\
&-&\frac{2A_+A_-\theta^2\omega^4}{\omega_{\theta}}[\sin(\delta_++\delta_-+2\omega_{\theta}T)-\sin(\delta_++\delta_-)]
\nonumber\\
&+&\frac{\theta\omega^2}{2}[A_+^2(e^{2\theta\omega^2T}-1)-A_-^2(e^{-2\theta\omega^2T}-1)]
\nonumber\\
&+&4A_+A_-\theta\omega^2T[\theta\omega^2\cos(\delta_+-\delta_-)+\omega_{\theta}\sin(\delta_--\delta_+)].
\end{eqnarray}

At this place we should emphasize the fact that for two types of solutions for Euler-Lagrange equations (35) two appropriate classical actions (39) and (47) are obtained. It is left for the further research to answer an open question which of these two are more suitable for the description of quantum dynamics of this noncommutative model.

\section{Wheeler-DeWitt Equation of the Model}

Within the standard quantum approach, i.e. on real and commutative spaces, following the procedure of canonical quantization, classical variables become quantum observables represented by Hermitian operators in the appropriate space of state. Then, the Hamiltonian constraint (15) becomes stationary Schr\"{o}dinger equation, so-called Wheeler-DeWitt equation. Incase of the Kantowski-Sachs model of the interior of the Schwarzschild black hole one gets:
\begin{equation}
\hat{\cal H}\Psi(u,v)=\left[-\frac{1}{4M_{pl}^2{V_0}}(\hat{\pi}_u^2-\hat{\pi}_v^2)-{V_0}M_{pl}^2(\hat{u}^2-\hat{v}^2)\right]\Psi(u,v)=0,
\end{equation}
but it can be presented also in the coordinate representation as:
\begin{equation}
\left[-\frac{\partial^2}{\partial{u}^2}+\frac{\partial^2}{\partial{v}^2}+\tilde{\omega}^2(u^2-v^2)\right]\Psi(u,v)=0,
\end{equation}
where $\tilde{\omega}=2V_0M_{pl}^2=\frac{V_0}{4\pi G}$.
This is the equation of a quantum isotropic oscillator-ghost-oscillator system with zero energy. By the method of separation of variables, and by inserting $\Psi_{n_1, n_2}(u,v)=\mu_{n_1}(u)\tau_{n_2}(v)$ into (49) one gets eigenstates:
\begin{eqnarray}
\Psi_{n_1, n_2}(u,v)&=&\mu_{n_1}(u)\tau_{n_2}(v)
\nonumber\\
&=&\left(\frac{\tilde{\omega}}{\pi}\right)^{\frac{1}{4}}\left[\frac{H_{n_1}(\sqrt{\tilde{\omega}}u)}{\sqrt{2^{n_1}n_1!}}\right]e^{-\frac{\tilde{\omega}{u^2}}{2}}
\left(\frac{\tilde{\omega}}{\pi}\right)^{\frac{1}{4}}\left[\frac{H_{n_2}(\sqrt{\tilde{\omega}}v)}{\sqrt{2^{n_2}n_2!}}\right]e^{-\frac{\tilde{\omega}{v^2}}{2}},
\end{eqnarray}
where $n_1,n_2\in N_0$ are quantum numbers of eigenstates of two decoupled oscillators. The eigenstates (50) satisfy the orthonormalization condition:
\begin{equation}
\int\!\!\!\int\Psi_{n_1, n_2}(u,v)\Psi_{m_1,m_2}(u,v)\,du\,dv=\delta_{n_1, m_1}\delta_{n_2, m_2},
\end{equation}
which follows from the normalization condition of the Hermite polynomials $H_n(x)$:
\begin{equation}
\int_{-\infty}^{\infty}e^{-x^2}H_{n}(x)H_{m}(x)\,dx=2^n\sqrt{\pi}n!\delta_{nm}.
\end{equation}

On the other hand, with $\Psi_{n_1,n_2}=\left|n_1,n_2\right\rangle$ from (48) one gets:
\begin{equation}
\hat{\cal H}\left|n_1,n_2\right\rangle=(n_1-n_2)\left|n_1,n_2\right\rangle=0,
\end{equation}
and consequently: $n_1=n_2=n$.

Then, the general solution of Wheeler-DeWitt equation (49) will be superposition of eigenstates (50) for all values of the quantum number $n_1=n_2=n\in N_0$:
\begin{equation}
\Psi(\!u,v\!)\!=\!\sum_{n=0}^{\infty}C_n\Psi_{n,n}(u,v)\!=\!\left(\frac{\tilde{\omega}}{\pi}\right)^{\frac{1}{2}}\!\sum_{n=0}^{\infty}\frac{C_n}{2^nn!}e^{-\frac{\tilde{\omega}}{2}(u^2+v^2)}H_n(\sqrt{\tilde{\omega}}u)H_n(\sqrt{\tilde{\omega}}v),
\end{equation}
which is interpreted, within a standard quantum approach, as a wave function for the Schwarzschild black hole interior presented through Kantowski-Sachs minisuperspace model.

\section{Adelic Wave Function of the Model}

The $p$-adic vacuum Kantowski-Sachs model has been considered in the chapter three and the conditions for existence of the vacuum $p$-adic states were determined. It is necessary condition for this model to be adelic one. Now we can construct the adelic wave function of the model $\Psi^{(adel.)}(u,v)$. This function is defined as an infinite product \cite{jpap{27},jpap{28}}:
\begin{eqnarray}
\Psi^{(adel.)}(u,v)=\Psi^{(\infty)}(u,v)\prod_{p\in M}{\Psi_p(u,v)}\prod_{p{\not\in M}}{\Omega(|u|_p)\Omega(|v|_p)},
\end{eqnarray}
where $\Psi^{(\infty)}(u,v)$ is a wave function in the real case represented by the expression (54), $\Psi_p(u,v)$ are $p$-adic wave functions which represent $p$-adic states of the model different from the vacuum states $\Omega(|u|_p)\Omega(|v|_p)$. The set $M$ is has to be finite one, otherwise the wave function $\Psi^{(adel.)}(u,v)$ would not belong to Hilbert space over adeles anymore.

The simplest vacuum adelic state of the model is:
\begin{equation}
\Psi_0^{(adel.)}(u,v)=\Psi_0^{(\infty)}(u,v)\prod_{p}{\Omega(|u|_p)\Omega(|v|_p)},
\end{equation}
where $\Psi_0^{(\infty)}(u,v)$ is the vacuum state in the real case, given in the first term in (54) i.e. for $n=0$.

At this moment we should emphasize that the interpretation of $p$-adic quantum mechanics results is done within formalism of adelic mechanics which, being more general, covers both $p$-adic and standard quantum mechanics. Namely, results of all measurements belong to the set of rational numbers $Q$. This means that theoretical results should be interpreted within this set as well. On the other hand, set $Q$ is common both for the field of $p$-adic numbers $Q_p$ and the field of real numbers $Q_{\infty}$. Becouse of this we consider minisuperspace coordinates to be rational numbers. Regarding the fact that $Q=Z\cup (Q\setminus Z)$, where $Z$ is the set of integers, coordinates $u$ and $v$ can have values either from $Z$ or $Q\setminus Z$.

Adopting the usual probability interpretation of the adelic wave function (55), we have:
\begin{eqnarray}
\left|\Psi^{(adel.)}(u,v)\right|_\infty^2=\left|\Psi^{(\infty)}(u,v)\right|_\infty^2\prod_{p\in M}{\left|\Psi_p(u,v)\right|_\infty^2}\prod_{p{\not\in M}}{\Omega(|u|_p)\Omega(|v|_p)},
\end{eqnarray}
because $\Omega^2(|u|_p)\Omega^2(|v|_p)\!=\!\Omega(|u|_p)\Omega(|v|_p)$. 

In case of the vacuum adelic state $\Psi_0^{(adel.)}(u,v)$ (56) one gets:
\begin{equation}
\left|\Psi_0^{(adel.)}(u,v)\right|_\infty^2=
\left\{ \begin{array}{ll}
\left|\Psi_0^{(\infty)}(u,v)\right|_\infty^2, & \mbox{$u,v\in Z$},\\
0, & \mbox{$u,v\in Q\setminus Z$}.
\end{array} \right.
\end{equation}
This result leads to some discretization of minisuperspace coordinates $u$ and $v$, because for all rational points density probability is non-zero only in the integer points of $u$ and $v$ (this discreetness is meantioned in introduction as a common characteristic of $p$-adic/adelic and noncommutative approach in quantum cosmology)\cite{jpap{13}}. Keeping in mind that $\Omega$ function is invariant with respect to the Fourier transform, this conclusion is also valid for the momentum space. Note that this kind of discreteness depends on adelic quantum state of the universe. When some $p$-adic states (for $p\in M$) are different from $\Omega (|u|_p)\Omega (|v|_p)$, then the above adelic discreteness becomes less transparent.

\section{Thermodynamics of the Model}

Until now we have seen that the consideration of the dynamics of the Schwarzschild black hole interior, in classical or quantum case, can be reduced to consideration of dynamics of classical or quantum system of two decoupled oscillators of equal frequencies, with zero energy in total. 

Starting with Hamiltonian (13), we will write Wheeler-DeWitt equation in the following form (in the International System of Units):

\begin{equation}
\left[-\frac{\tilde{N}}{4{V_0}M_{pl}^2}(\hat{\pi}_u^2-\hat{\pi}_v^2)\frac{\hbar}{c^2}-\tilde{N}{V_0}M_{pl}^2(\hat{u}^2-\hat{v}^2)\frac{c^2}{\hbar}\right]\Psi(u,v)=0,
\end{equation}
actually in coordinate representation:
\begin{equation}
\left[-\frac{\tilde{N}}{4{V_0}M_{pl}^2}(-{\hbar}^2\frac{\partial^2}{\partial{u}^2}+{\hbar}^2\frac{\partial^2}{\partial{v}^2})\frac{\hbar}{c^2}-\tilde{N}{V_0}M_{pl}^2(u^2-v^2)\frac{c^2}{\hbar}\right]\Psi(u,v)=0.
\end{equation}
With $\Psi_{n_1, n_2}(u,v)=\mu_{n_1}(u)\tau_{n_2}(v)$, (60) becomes decoupled with respect to $u$ and $v$:
\begin{equation}
\left[-\frac{{\hbar}^3\tilde{N}}{4{V_0}c^2M_{pl}^2}\frac{\partial^2}{\partial{u}^2}+\frac{\tilde{N}V_0c^2M_{pl}^2}{\hbar}u^2\right]\mu_{n_1}(u)=E_{n_1}\mu(u),
\end{equation}
\begin{equation}
\left[-\frac{{\hbar}^3\tilde{N}}{4{V_0}c^2M_{pl}^2}\frac{\partial^2}{\partial{v}^2}+\frac{\tilde{N}V_0c^2M_{pl}^2}{\hbar}v^2\right]\tau_{n_2}(v)=E_{n_2}\tau(v),
\end{equation}
and there are the equations of two decoupled linear harmonic oscillators (where, according to (53), $n_1=n_2=n$ and then $E_{n_1}=E_{n_2}=E_n$). The term $\frac{\tilde{N}V_0c^2M_{pl}^2}{\hbar}u^2$ in (61), actually $\frac{\tilde{N}V_0c^2M_{pl}^2}{\hbar}v^2$ in (62), matches the potential energy $E_p(u)$ of oscillator by coordinate $u$, i.e. potential energy $E_p(v)$ of oscillator by coordinate $v$, respectively. Then, the frequency of the oscillator will be:
\begin{equation}
\omega^2=\frac{1}{\tilde{m}}\frac{\partial^2E_p(u)}{\partial{u^2}}=\frac{1}{\tilde{m}}\frac{\partial^2E_p(v)}{\partial{v^2}}=\frac{\tilde{N}}{\tilde{m}}\frac{2V_0c^2M_{pl}^2}{\hbar},
\end{equation}
where $\tilde{m}$ is the mass of the oscillator.

Thereby, we should point at the possibility of interpretation of these oscillators as two gravitational degrees of freedom. One of them is related to interior of a black hole with the energy $E_n$, and the other degree of freedom is related to interior of the corresponding, sometimes referred to, as a "white hole" with the energy $-E_n$. This interpretation offers the possibility that by applying Feynman-Hibbs procedure \cite{autbk2} to a part of Wheeler-DeVitt equation that is related to one oscillator, the temperature could be defined as well as entropy of a black hole with appropriate quantum corrections.

The freedom of a choice of the lapse function $\tilde{N}$ enables us to fix the gauge $\frac{\tilde{N}}{\tilde{m}}=\frac{6c^2}{V_0\hbar}$ so that we could obtain the correct expression for Hawking temperature of a black hole (the Barbero-Immirzi parameter is fixed in a similar manner in loop quantum gravity \cite{jpap{29}}). Then from (63) we obtain following:
\begin{equation}
\hbar\omega=\sqrt{\frac{3}{2\pi}}E_{pl},
\end{equation}
where $E_{pl}=\sqrt{\frac{\hbar{c^5}}{G}}\approx 1.22\times10^{19}$ GeV is the (unreduced) Planck energy.

The classical partition function for the linear harmonic oscillator (LHO) of mass $m$ and frequency $\omega$ is:
\begin{equation}
Z_{class}=\frac{1}{\beta\hbar\omega}.
\end{equation}
where $\beta=\frac{1}{kT}$, where $T$ is the temperature of the system and $k$ is the Boltzmann constant. The "corrected" potential for the case of quantum LHO \cite{autbk2}:
\begin{equation}
V=\frac{m\omega^2}{2}\left(x^2+\frac{\beta\hbar^2}{12m}\right),
\end{equation}
gives partition function:
\begin{equation}
Z_{approx}=\frac{e^{-\frac{(\beta\hbar\omega)^2}{24}}}{\beta\hbar\omega}.
\end{equation}
By substituting (64) into (67) one gets that in the quantum case for this model of the black hole:
\begin{equation}
Z_{approx}=\sqrt{\frac{2\pi}{3}}\frac{e^{-\frac{\beta^2E_{pl}^2}{16\pi}}}{\beta E_{pl}}.
\end{equation}
Then, the internal energy of the black hole is:
\begin{equation}
\bar{E}=-\frac{\partial}{\partial\beta}(\ln Z_{approx})=\frac{E_{pl}^2}{8\pi}\beta+\frac{1}{\beta}=mc^2,
\end{equation}
where $m$ is the mass of the black hole. By solving (69) for $\beta$, with $E_{pl}\ll mc^2$, one gets:
\begin{equation}
\beta=\frac{8\pi{mc^2}}{E_{pl}^2}\left[1-\frac{1}{8\pi}\left(\frac{E_{pl}}{mc^2}\right)^2\right],
\end{equation}
actually, in terms of the Hawking temperature $\beta_H=\frac{8\pi{mc^2}}{E_{pl}^2}=\frac{1}{kT_H}$, temperature of the black hole with quantum correction will be:
\begin{equation}
\beta=\beta_H\left[1-\frac{1}{\beta_H{mc^2}}\right].
\end{equation}

The entropy of the model with quantum correction one gets from:
\begin{equation}
\frac{S}{k}=\ln Z_{approx}+\beta\bar{E}.
\end{equation}
By supstituting (67), (69) and (70) into (72) one gets:
\begin{equation}
\frac{S}{k}=\frac{A_s}{4l_{pl}^2}\left[1\!-\!\frac{1}{8\pi}\left(\frac{E_{pl}}{mc^2}\right)^2\right]\!-\!\frac{1}{2}\ln\left(\frac{A_s}{4l_{pl}^2}\left[1\!-\!\frac{1}{8\pi}\left(\frac{E_{pl}}{mc^2}\right)^2\right]^2\right)\!-\!\frac{1}{2}\ln(24)\!+\!1,
\end{equation}
where $A_s=4\pi{r_g}^2$ is the surface of the event horizon of the black hole whose Schwarzschild radius is $r_g=\frac{2Gm}{c^2}$, while $l_{pl}=\sqrt{\frac{\hbar{G}}{c^3}}\approx 1.62\times10^{-35}$ m is the Planck length.
In terms of the Bekenstein-Hawking entropy $\frac{S_{BH}}{k}=\frac{A_s}{4l_{pl}^2}$, with $E_{pl}\ll mc^2$, (73) becomes:
\begin{equation}
\frac{S}{k}=\frac{S_{BH}}{k}-\frac{1}{2}\ln\left(\frac{S_{BH}}{k}\right)+{\cal O}(S_{BH}^{-1}),
\end{equation}
which is exsspresion for the entropy of the black hole with quantum correction, that was obtained in a different manner in papres Ref. \cite{jpap{30},jpap{31},jpap{32},jpap{33},jpap{34}}, as well, and it corresponds to what is obtained in the string theory \cite{jpap{35}} and in loop quantum gravity \cite{jpap{36},jpap{37},jpap{38},jpap{39},jpap{40}}.

\section{Conclusion}

In this paper we presented the dynamics of the interior of the non-rotating and non-charged, Schwarzschild, black hole, as the dynamics of the homogeneous and anisotropic Kantowski-Sachs minisuperspace two-oscillator cosmological model. We determined classical action and Feynman propagators in a $p$-adic and noncommutative case. Especially, for the $p$-adic model, the conditions for existence of the vacuum $p$-adic states are determined. In the noncommutative case classical actions for two types of solutions for Euler-Lagrange equations are calculated. Then, within standard commutative cosmology, Wheeler-DeWitt equation of the model is written and its solution, i.e. the wave function of the model, is determined and by means of that, quantum dynamics of the interior of the Schwarzschild black hole is described. Within the adelic approach, the adelic wave function is written as a product of real and $p$-adic wave functions of the model. Finally, in the part which is related to the Schwarzschild black hole thermodynamics, by applying the Feynman-Hibbs procedure, Hawking temperature is determined as well as entropy with corresponding quantum corrections.

The important question for of this further research is the question of the classical limit i.e. prediction of classical state of the model from its wave function. In that sense it is of the great interest to consider  classical-quantum correspondence through classical trajectory in configuration space of variables $u$ and $v$ (which we obtain from (19) by eliminating $t$) and maximum of square modulus of the wave function from (54)
\cite{jpap{21}}.

%
%

\begin{acknowledgements}
The work of G. Djordjevic and Lj. Nesic is partially supported by Ministry of Education, Science and Technological Development of the Republic of Serbia under Grant Nos. 174020 and 176021, and ICTP-SEENET-MTP Project PRJ09 “Cosmology and Strings” in the frame of the Southeastern European Network in Theoretical and
Mathematical Physics, the work of D. Radovancevic is partially supported within this frame, as well. G. Djordjevic is thankful to the CERN-TH group for financial support and hospitality during his stay, where a part of this paper was finalized.
\end{acknowledgements}

\appendix

\section*{\\Appendix A. Coefficients $\alpha_{ij}$ i $K_i^\pm$}

If $A=\frac{2\theta\omega^2\Omega_1}{\omega_\theta^2-\Omega_1^2}$, $B=\frac{2\theta\omega^2\Omega_2}{\omega_\theta^2-\Omega_2^2}$ and\\ $\Delta=-2AB[1-\cos(\Omega_1T)\cos(\Omega_2T)]+(A^2+B^2)\sin(\Omega_1T)\sin(\Omega_2T)\not=0$, then:
\begin{eqnarray}
\alpha_{11}&=&\frac{1}{\Delta}[-AB+B^2\sin(\Omega_1T)\sin(\Omega_2T)+AB\cos(\Omega_1T)\cos(\Omega_2T)],
\nonumber\\
\alpha_{12}&=&\frac{1}{\Delta}[AB(\cos(\Omega_2T)-\cos(\Omega_1T))],
\nonumber\\
\alpha_{13}&=&\frac{1}{\Delta}[B\sin(\Omega_1T)\cos(\Omega_2T)-A\sin(\Omega_2T)\cos(\Omega_1T)],
\nonumber\\
\alpha_{14}&=&\frac{1}{\Delta}[A\sin(\Omega_2T)-B\sin(\Omega_1T)],
\nonumber\\
\alpha_{21}&=&\frac{1}{\Delta}[AB\sin(\Omega_1T)\cos(\Omega_2T)-B^2\cos(\Omega_1T)\sin(\Omega_2T)],
\nonumber\\
\alpha_{22}&=&\frac{1}{\Delta}[B^2\sin(\Omega_2T)-AB\sin(\Omega_1T))],
\nonumber\\
\alpha_{23}&=&\frac{1}{\Delta}[B-B\cos(\Omega_1T)\cos(\Omega_2T)-A\sin(\Omega_1T)\sin(\Omega_2T)],
\nonumber\\
\alpha_{24}&=&\frac{1}{\Delta}[B(\cos(\Omega_1T)-\cos(\Omega_2T))],
\nonumber\\
\alpha_{31}&=&\frac{1}{\Delta}[-AB+AB\cos(\Omega_1T)\cos(\Omega_2T)+A^2\sin(\Omega_1T)\sin(\Omega_2T)],
\nonumber\\
\alpha_{32}&=&\frac{1}{\Delta}[AB(\cos(\Omega_1T)-\cos(\Omega_2T))],
\nonumber\\
\alpha_{33}&=&\frac{1}{\Delta}[A\sin(\Omega_2T)\cos(\Omega_1T)-B\sin(\Omega_1T)\cos(\Omega_2T)],
\nonumber\\
\alpha_{34}&=&\frac{1}{\Delta}[A\sin(\Omega_2T)-B\sin(\Omega_1T)],
\nonumber\\
\alpha_{41}&=&\frac{1}{\Delta}[AB\cos(\Omega_1T)\sin(\Omega_2T)-A^2\sin(\Omega_1T)\cos(\Omega_2T)],
\nonumber\\
\alpha_{42}&=&\frac{1}{\Delta}[A^2\sin(\Omega_1T)-AB\sin(\Omega_2T)],
\nonumber\\
\alpha_{43}&=&\frac{1}{\Delta}[A-B\sin(\Omega_1T)\sin(\Omega_2T)-A\cos(\Omega_1T)\cos(\Omega_2T)],
\nonumber\\
\alpha_{44}&=&\frac{1}{\Delta}[A(\cos(\Omega_2T)-\cos(\Omega_1T))].
\end{eqnarray}
\begin{eqnarray}
K_1^{\pm}&\!=\!&K_1^{\pm}(\!\Omega_1\!)\!\!=\!(\!1{\mp}A^2\!)[(\!1+\theta^2\omega^2\!)\Omega_1^2{\mp}\omega^2]\!-2\theta\Omega_1A(\omega^2\pm\omega^2),
\nonumber\\
K_2^{\pm}&\!=\!&K_2^{\pm}(\!\Omega_2\!)\!\!=\!(\!1{\mp}B^2\!)[(\!1+\theta^2\omega^2\!)\Omega_2^2{\mp}\omega^2]\!-2\theta\Omega_2B(\omega^2\pm\omega^2),
\nonumber\\
K_3^{\pm}&\!=\!&K_3^{\pm}(\!\Omega_1,\Omega_2\!)\!\!=\![(\!1+\theta^2\omega^2)\Omega_1\Omega_2{\mp}\omega^2](\!1{\mp}AB\!){\mp}(\!A{\pm}B\!)\theta\omega^2\!(\!\Omega_1{\pm}\Omega_2\!).
\end{eqnarray}
\newpage



\end{document}